# Stable Spin State Analysis of Fe, Co, Ni-modified Graphene-ribbon

Norio Ota

Graduate School of Pure and Applied Sciences, University of Tsukuba, *1-1-1 Tenoudai Tsukuba-city 305-8571, Japan*

Magnetic graphene-ribbon is a candidate for realizing future ultra high density 100 tera bit/inch$^2$ class data storage media. In order to increase the saturation magnetization, first principles DFT analysis was done for Fe, Co, Ni-modified zigzag edge graphene-ribbon. Typical unit cell is [C32H2Fe1], [C32H2Co1] and [C32H2Ni1] respectively. Most stable spin state was Sz=4/2 for Fe-modified case, whereas Sz=3/2 for Co-case and Sz=2/2 for Ni-case. Magnetic moment of Fe,Co, and Ni were 3.63, 2.49 and 1.26 μ$_B$, which can be explained by the Hund-rule considering charge donation to neighboring carbons. Band calculation shows half-metal like structure with a large band gap (in Co-case, 0.55eV) for up-spin, whereas very small gap (0.05eV) for down-spin, which will be useful for many featured application like information storage, spin filter and magneto-resistance devices. Dual layer Fe-modified ribbon shows a tube like curved structure, which may suggest a carbon nanotube creation by Fe catalyst.

## 1. Introduction

Current magnetic data storage[1)-2)] has a density around 1 tera-bit/inch$^2$ with 10 nm length, 25 nm width magnetic bit. We need a future ultra high density 100 tera bit /inch$^2$ magnetic data storage materials. One promising candidate is a ferromagnetic molecule size dot array having a typical areal bit size of 1 nm by 2.5 nm as illustrated in Fig.1. Recently, carbon based room-temperature ferromagnetic materials are experimentally reported[4)-9)]. They are graphite and graphene like materials. From a theoretical view point, Kusakabe and Maruyama[10)-11)] proposed an asymmetric graphene-ribbon model with two hydrogen modified (dihydrogenated) zigzag edge carbon showing ferromagnetic behavior. Our previous papers[12)-14)] have reported multiple spin state analysis of graphene like molecules, which suggest a capability of strong magnetism. Especially, in our recent analysis[15)], bare carbon zigzag edge graphene-ribbon shows a possibility of strong magnetism with nano-meter width and long straight line looks like a recording track. However, we need more larger magnetization graphene-ribbon.

Here, by using the first principles density functional theory (DFT) based analysis, ferromagnetic atomic species like Fe, Co and Ni were tried to modify zigzag edge carbon of graphene-ribbon to find a capability of strong magnetism. Typical unit cell were [C32H2Fe1], [C32H2Co1] and [C32H2Ni1] respectively. Additionally, dual layer Fe-modified ribbon were analyzed for studying a capability of multilayer magnetization enhancement.

Already, there were some experiments of graphene formation on Fe(110) substrate[16)], and Co layer structure observation on graphite substrate[17)]. In DFT calculation, iron cluster on graphene sheet[18)], or iron –based molecule grafted on graphene[19)] were studied. Those were the cases that iron or cobalt atoms coupled with graphene surface sheet. In viewpoint of information storage and/or data processing devices, narrow stripe straight line like a graphene-ribbon is very suitable for industrial application. Therefore, this report focuses on graphene-ribbon and its chemical modification by magnetic species.

## 2. Model graphene-ribbon

Bird eye view of typical graphene-ribbon model is shown in Fig.2. Modified Fe atom bonds with zigzag edge positioned two carbons of one side (left side in the figure) of graphene-ribbon. Whereas, another side (right side) zigzag edges are all hydrogenated. Iron (Fe) has two 4s-orbital electrons and reasonably bond with sigma-electrons of two carbons. As illustrated in Fig.2, track width is 1.8nm. In tracking direction, length of 1nm includes five zigzag carbon edges, which fits a size for 100 tera bit /inch$^2$ areal density. In Fig.3, one periodic unit cell for an infinite length graphene-ribbon is shown in a red square mark as [C32H2Fe1], where blue ball show Fe, gray ball carbon and small ball hydrogen, also in (b), [C32H2Co1] and in (c), [C32H2Ni1].

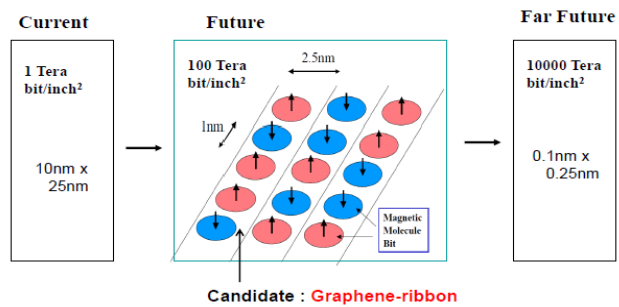

**Fig.1** Future ultra-high density 100 tera-bit/inch$^2$ magnetic data storage with a bit size of 1 nm by 2.5nm. Graphene-ribbon is one candidate for such application.

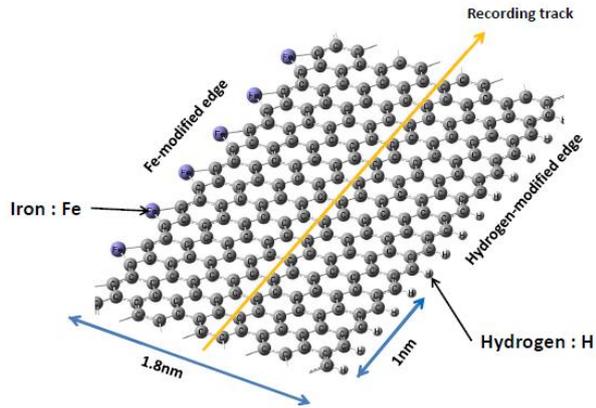

**Fig.2** Bird eye view of Fe-modified graphene-ribbon. One iron atom bonds with two zigzag edge carbons at left side, whereas right side edge carbons are all hydrogenated.

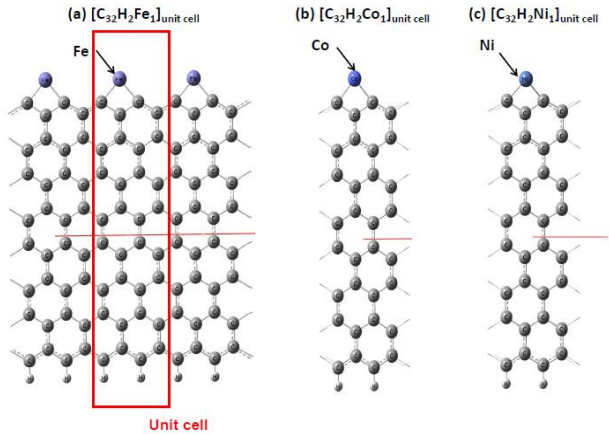

**Fig.3** Model graphene-ribbon of $[C_{32}H_2Fe_1]_{unit\ cell}$ in (a), where blue ball show Fe (or Co, Ni), gray ball carbon and small ball hydrogen, also in (b),[C32H2Co1] and in (c),[C32H2Ni1].

### 3. Calculation Method

We have to obtain the (1) optimized atom configuration, (2) total energy, (3) spin density configuration, (4) magnetic moment of every atom and (5) band characteristics depending on a respective given spin state Sz to clarify magnetism. Density functional theory (DFT) [20)-21)] based generalized gradient approximation (GGA-PBEPBE) [22)] was applied utilizing Gaussian03 package[23)] with an atomic orbital 6-31G basis set[24)]. In this paper, total charge of model unit-cell is set to be completely zero. In this unrestricted DFT calculation, S(S+1) value is obtained to check a degree of spin-contamination. Inside of a unit-cell, three dimensional DFT calculation was done. One dimensional periodic boundary condition was applied to realize an unlimited length graphene-ribbon. Self-consistent energy, atomic configuration and spin density calculations are repeated until to meet convergence criteria. The required convergence on the root mean square density matrix was less than 10E-8 within 128 cycles.

### 4, Stable spin state

Typical graphene-ribbon model is shown in Fig.3. Each unit-cell has limited numbers of unpaired electrons, which enable allowable numbers of multiple spin states. For example, in case of [C32H2Fe1], there are five spin state like Sz=8/2, 6/2, 4/2, 2/2 and 0/2. Starting DFT calculation, one certain Sz value should be installed as a spin parameter. After that, all of the self-consistent calculations were repeated until to meet a convergence criteria.

Optimized atomic configuration resulted a flat and straight ribbon. Some theoretical papers suggested that narrow width ribbon less than 1.5nm may show some twisting and uniform curving[25)-26)]. However, our model has sufficient large width (1.8nm) not occurring such irregularity. Ribbon became a flat and straight configuration. Distance between Fe-C (zigzag edge) was 0.187nm for Sz=4/2, Co-C 0.184nm for Sz=3/2 and Ni-C 0.183nm for Sz=2/2, which are reasonable with atomic radius.

The first question is which spin state is the most stable one. Converged energy for given Sz were compared as shown in Fig.4, where the most lowest energy was defined as zero energy to simply compare energy differences between spin states. In case of [C32H2Fe1] (blue square mark), Sz=4/2 and 2/2 were lowest spin states. Again, which is the most stable one? In magnetic calculation, we should take care spin-contamination[29)30)], which occurs from unsuitable spin configuration. We can compare spin contamination by the difference between DFT obtained S(S+1) and installed Sz(Sz+1) values. In case of Sz=4/2, S(S+1) was 6.40, whereas Sz(Sz+1) was 6.00, which were close together and suggested that this spin state gives less spin-contamination. Whereas, in case of Sz=2/2, S(S+1) was 3.40 compared with Sz(Sz+1)=2.00, which was fairly large spin-contamination. We should select the most stable spin state to be Sz=4/2.

Similar phenomenon occurs in [C32H2Co], there are two lowest spin state Sz=3/2 and 1/2. Considering spin-contamination, Sz=3/2 was the most stable one. In case of [C32H2Ni1], stable spin state was Sz=2/2.

One typical spin density configuration was illustrated in Fig.5 in case of Sz=4/2 of [C32H2Fe1]. Red cloud shows up-spin, while blue cloud down-spin. Contour lines for 0.001, 0.1, 0.4e/A³ are arrowed. Iron atom wear very large up-spin cloud. Zigzag edge positioned carbon show down-spin, whereas next neighbor carbon up-spin. Inside of graphene-ribbon, carbon sites alternately arrange up- and down-spins one by one very regularly. This is a specific feature of pai-electron oriented spin configuration in graphene[15].

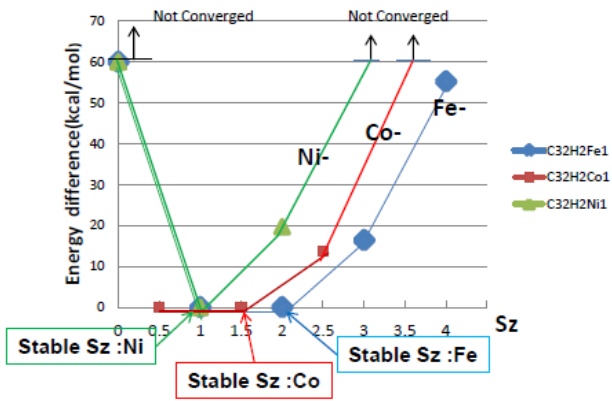

**Fig.4** Energy difference between spin state Sz for [C32H2Fe1] (blue square mark), [C32H2Co1] (red square) and [C32H2Ni1] (green triangle).

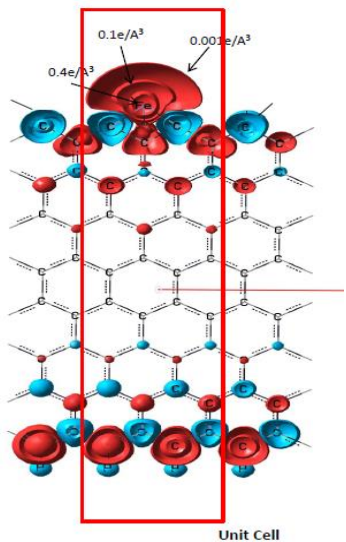

**Fig.5** Spin density configuration in [C32H2Fe1] graphene-ribbon. Red cloud shows up-spin, whereas blue one down-spin. Fe site has a very large up-spin density. Inside of graphene, up and down spin arranges one by one very regularly.

### 5, Magnetic moment and Hund-rule

Our purpose is to obtain a larger magnetization graphene-ribbon by Fe, Co and Ni modification. DFT calculation gives an atomic magnetic moment M. For [C32H2Fe1], Sz=4/2 case, Fe site has M(Fe)=3.63 $\mu_B$ as illustrated in Fig.6. We tried to explain this value based on the Hund-rule. Neutral Fe has six 3d-electrons, among them five are up-spins and rest one down-spin. Thus, simple Hund-rule gives $4\mu_B$. However, there is some amount of escaped charge from Fe. Charge of Fe is +0.5e, whereas summed charge of neighbor three carbons are -0.43e (-0.13, -0.13, -0.17e) as shown in Fig.6, which suggests that Fe donates extra charge 0.5e to those carbons. This extra charge has to be subtracted from 3d up-spin as shown in left hand side of Fig.6 in order to reduce the total exchange coupling energy between Fe and carbons. We applied this simple assumption to obtain magnetization M*, that is, M*(Fe)=(5.0 - 0.5)+(-1.0)=3.50 $\mu_B$, which value is close to DFT result of M(Fe)=3.63 $\mu_B$.

In Table1, such Hund-rule based estimated M* and DFT obtained M are summarized. In Co-case, M*(Co)=2.56 $\mu_B$ is close to M(Co)=2.49 $\mu_B$. Whereas, in Ni-case, M*(Ni)=1.62 $\mu_B$ shows somewhat discrepancy with M(Ni)=1.26 $\mu_B$.

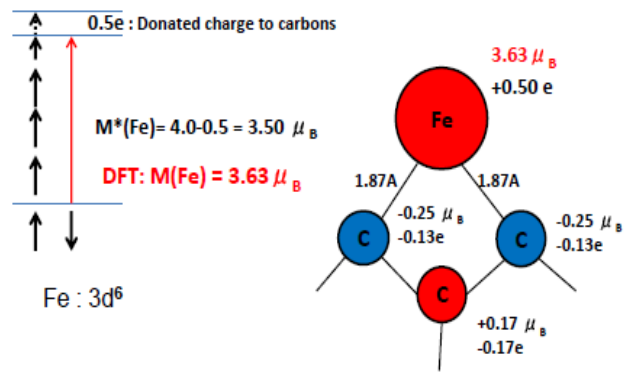

**Fig.6** Hund-rule based magnetic moment is estimated to be M*(Fe)=3.50, whereas DFT calculated one M(Fe)=3.63 $\mu_B$. Fe atom donates extra charge 0.5e to neighbor carbons. This charge should be subtracted from Fe 3d-electron up-spin site.

**Table 1** Stable spin state Sz for every unit cell. Magnetic moment M of Fe, Co, and Ni obtained by DFT calculation were compared with the Hund-rule based estimated magnetic moment M*.

|  | C32H2Fe1 | C32H2Co1 | C32H2Ni1 |
|---|---|---|---|
| Stable Sz | 4/2 | 3/2 | 2/2 |
| (a) Hund-rule magnetic moment $M_H$ ($\mu_B$) | 4 | 3 | 2 |
| (b) DFT calculated magnetic moment M ($\mu_B$) | 3.63 | 2.49 | 1.26 |
| (c) DFT Calculated charge (e) | 0.50 | 0.44 | 0.38 |
| (d) Estimated magnetic moment M* ($\mu_B$) | 3.50 | 2.56 | 1.62 |

### 6, Band analysis and energy gap

Graphene-ribbon is a periodic system with one-dimensional crystallography. Here, Band characteristics were analyzed. In case of [C32H2Fe1] unit cell, lattice parameter "a" is 0.507nm. We divided k-space to 12 elements from k=0/a to π/a.

Calculated band structure was illustrated in Fig.7. Red curves are up-spin occupied band (occupied crystal orbit), light red up-spin unoccupied band (virtual crystal orbit), blue down-spin occupied band and light blue down-spin unoccupied band.

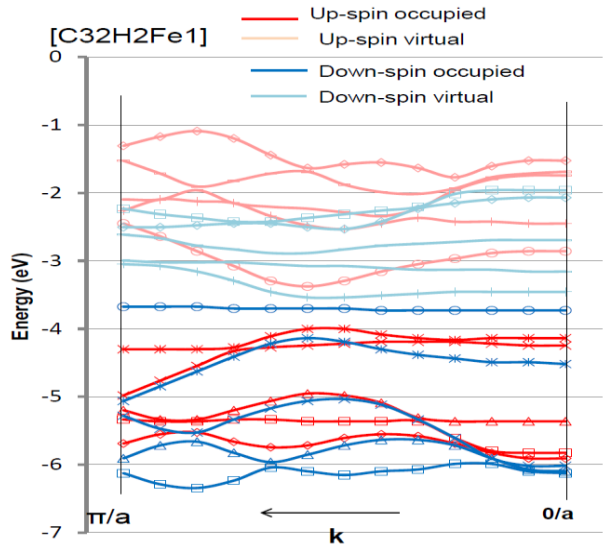

**Fig.7** Band characteristic of [C32H2Fe1]$_{unit\ cell}$.
Up-spin energy gap is 0.63eV. Among this gap, there are two down-spin orbits, which show almost half metallic band structure.

Band characteristic is very unique. Up-spin energy gap is 0.63eV. Among this gap, there are two down-spin orbits (one occupied and one unoccupied). Down-spin band looks almost metallic like. This suggests a capability of down-spin dominated half metal. Especially in case of Co-case as summarized in Table2, down-spin energy gap of [C32H2Co1] is only 0.05eV. Occupied up-spin orbits originate from Fe 3d-electrons, whereas down-spin from pai-conjugated orbits of graphene carbons. Graphene carbon part may be conductive for down-spin electrons. On the other hand, Fe, Co and Ni side part is resistive. Suggested by such distinct characteristics, we can expect future application as like a data storage, spin filter, and a large spin polarized magneto-resistance devices.

### 7. Multilayer ribbon

In order to increase the areal saturation magnetization per bit for information storage application, a simple way is to multiple layer numbers[15]. Here, we tried dual layer Fe-modified graphene-ribbon.

**Table 2** Band gap analysis of Fe, Co and Ni-modified graphene-ribbon.

|  | [C32H2Fe1] | [C32H2Co1] | [C32H2Ni1] |
|---|---|---|---|
| Up-spin HOCO (eV) | -3.99 | -4.04 | -4.16 |
| Up-spin LUCO (eV) | -3.36 | -3.49 | -3.51 |
| Up-spin indirect gap (eV) | 0.63 | 0.55 | 0.64 |
| Down-spin HOCO (eV) | -3.67 | -3.83 | -4.27 |
| Down-spin LUCO (eV) | -3.54 | -3.77 | -4.08 |
| Down-spin indirect gap (eV) | 0.12 | 0.05 | 0.19 |

HOCO; Highest occupied crystal orbit
LUCO; Lowest unoccupied crystal orbit

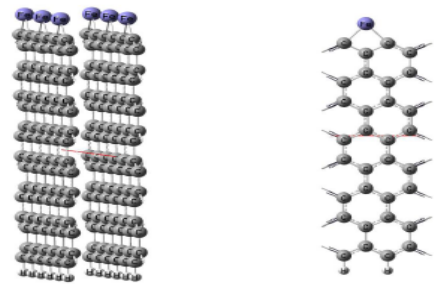

Bird eye view    Plane view
(Unit Cell)

**Fig.8** Dual layer [C32H2Fe1]x2 graphene-ribbon model. Bird eye view of initial setting of two ribbons and an overlapped plane view.

Initial setting of two ribbons is shown in Fig.8 as a bird eye view and an overlapped plane view. Initial distance between two graphene-ribbons was 0.355nm similar with a 3D-graphite layer to layer distance. Side view is shown in Fig.9 (a), Also, initial spin density side view is illustrated in (b), where spin clouds were mirror images each other. Installed spin state was Sz=8/2.

During costly atomic configuration calculation, parallel atomic configuration gradually deformed to necklace like unique shape as shown in (c). Distance from Fe to Fe became close 0.268nm. On the contrary, center part curved and expanded to 0.555nm. At the hydrogen end, atoms became close again to only 0.167nm. Converged spin density map was shown in (d). Both iron spins became up-spins and coupled each other. On the other hand, there was not any spin density configuration on lower graphene part including hydrogen. In several experiments[27)28)], there observed that iron particle and substrate act as catalyst to form carbon nanotube from graphite and graphene. Our calculation suggested that iron to iron exchange interaction plays an important role for those catalysis.

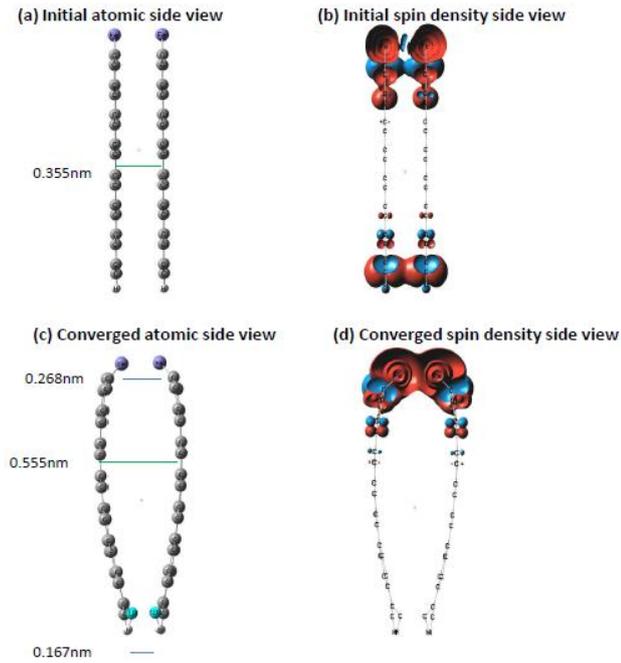

**Fig.9** Side view of dual layer Fe-modified graphene-ribbon. Initial configuration was parallel as shown in (a) and (b). Converged configuration was necklace like figure as (c) and (d).

## 8. Conclusion

For realizing a future 100 tera-bit/inch$^2$ class magnetic information storage, graphene-ribbon is very attractive. In order to increase the saturation magnetization, first principles DFT analysis was done for Fe, Co, Ni-modified zigzag edge graphene-ribbon. Typical unit cell is [C32H2Fe1], [C32H2Co1] and [C32H2Ni1] respectively. Conclusions are,

(1) Most stable spin state was $S_z=4/2$ for Fe-modified graphene ribbon, whereas $S_z=3/2$ for Co-case and $S_z=2/2$ for Ni-case.
(2) Atomic magnetization of Fe, Co and Ni were 3.63, 2.49 and 1.26 $\mu_B$, which can be explained by the Hund-rule based assumption regarding a charge donation to neighboring carbons as like 3.50, 2.56, and 1.62$\mu_B$.
(3) Band calculation shows half-metal like structure with relatively large band gap (0.5~0.6eV) for up-spin, whereas small gap (0.05~0.2eV) for down-spin. This suggests a capability of magnetic storage, spin-filter and large magneto-resistance devices.
(4) For multiplying the magnetization, dual layer Fe-modified ribbon was optimized. Calculation result shows a tube like curving structure, which may suggest a carbon nanotube creation by Fe catalyst.

## Acknowledgment

I would like to say great thanks to Prof. Takeshi Inoshita for triggering me this study.